\documentclass[twocolumn,showpacs,preprintnumbers,amsmath,amssymb]{revtex4}

\usepackage{epsf}
\usepackage{rotating}

\usepackage{graphicx,epsfig}% Include figure files
\usepackage{dcolumn}% Align table columns on decimal point
\usepackage{bm}% bold math
\def\cA{{\cal A}}

\newcommand{\req}[1]{Eq.~(\ref{#1})}
\newcommand{\avg}[1]{\langle #1\rangle}

\def\puk{p_{u,k}}

\begin{document}

\preprint{}

\title[Title]
{Phase Transitions in Transportation Networks with Nonlinearities}

\author{C.~H.~Yeung and K.~Y.~Michael Wong}
\affiliation{Department of Physics, The Hong Kong University
of Science and Technology, Hong Kong, China}

\date{\today}

\begin{abstract}
We investigate a model of transportation networks with nonlinear elements 
which may represent local shortage of resources.
Frustrations arise from competition for resources. 
When the initial resources are uniform,
different regimes with discrete fractions of satisfied nodes are observed,
resembling the Devil's staircase.
We demonstrate how functional recursions are converted to simple 
recursions of probabilities.
Behavior similar to those in the vertex cover 
or close packing problems are found.
When the initial resources are bimodally distributed, 
increases in the fraction of rich nodes induce a glassy transition, 
entering an algorithmically hard regime.
\end{abstract}
% 02.50.-r: Probability theory, stochastic processes, and statistics
% 05.20.-y: Classical statistical mechanics
% 89.20.-a: Interdisciplinary applications of physics

\maketitle

%%%%%%%%%%%%%%%%%%%%%%%%%%%%%%%%%%%%%%%%%%%%%%%%%%%%%%%%%%%%%%%%%%%%%%%%%%

The study of currents and flows in networks 
is one of the most important problems 
in physics and many other areas of application~\cite{doyle1984}. 
Besides electric circuits transporting electric currents, 
there are transportation networks, communications networks, 
hydraulic networks, mammalian circulatory systems 
and vascular systems in plants~\cite{rockafellar1984,banavar2000}. 
A unified approach to these problems 
is facilitated by the minimization of cost functions. 
For example, they may represent the dissipation energy 
(via Thomson's principle for electric currents)~\cite{doyle1984} 
or time delays in communications networks. 
There is a close relation 
between the flow patterns and the cost functions. 
For example, it was found that the flow pattern is tree-like 
when the cost function is convex, 
but loopy otherwise~\cite{banavar2000}. 
These cost functions are continuous 
and often describe resistive flows. 
On the other hand, networks may contain nonlinear elements
such as step-like penalties 
that may affect the flow distribution. 

The inclusion of nonlinear elements 
can drastically modify the flow patterns in the network. 
Nonlinearities are often represented by cost functions 
that are discontinuous in the currents. 
This creates numerous choices 
in deciding the current-carrying links and the idle ones. 
As shown in this paper, 
the flow patterns can demarcate clusters of nodes 
{\it receiving} currents. 
Clusters formed by similar energetic considerations 
have been found to play an important role in disordered systems 
such as the random field Ising model (RFIM)~\cite{imry1975}, 
giving rise to the so-called Griffiths singularities 
and cascades of phase transitions~\cite{bruinsma1983}. 
As we shall see, similar behavior can also be found 
in nonlinear networks.

The origin of these interesting phenomena 
can be traced the presence of {\it frustrations}, 
which refers to the conflict bewteen competing interaction energies 
in the system~\cite{toulouse1977}. 
This connects such transportation networks 
with a broad class of network systems 
in which frustration is inherent, 
including the $K$-satisfiability~\cite{mezard2002},
the graph coloring problem~\cite{mulet2002}, 
and error-correcting codes~\cite{nishimori2001}.
Spin glasses, the prototype of frustrated systems, 
provide the statistical mechanics to study these systems~\cite{mezard1987}.

Transportation networks 
consist of nodes with either surplus or shortage of resources, 
and an important problem is to distribute them
to achieve a networkwide satisfaction with 
a minimum transportation cost~\cite{wong2006}.
This problem is important in load balancing in computer networks 
and network flow of commodities~\cite{shenker1996}.
The focus of this paper is on the case 
where optimization of resources can proceed by penalizing nodes with shortages.
In applications such as communications networks, 
such shortages can be detrimental to the performance. 
Hence it is interesting to consider step-like shortage costs, 
which give rise to
unique behavior and physical picture 
absent in the previous models~\cite{wong2006}.
Frustrations arise from competition for resources among connected nodes.
Numerous metastable states emerge,
leading to typical glassy behavior. 

Many NP-complete problems 
in computational complexity theory~\cite{garey1979}
exhibit glassy behavior.
As we shall see,
the main results in this paper 
are that changes in the shortage per node 
induce phase transitions to glassy behavior. 
A new feature in our model 
is a cascade of phase transitions 
resembling the Devil's staircase in the glassy phase, 
and the first cascade 
is similar to the vertex cover problem~\cite{weigt2000}. 
As a crucial difference from other NP-complete problems,
the present problem involve continuous variables 
which complicate the analyses.
We demonstrate how recursions of functions with continuous variables
can be converted to simple recursions of probabilities.

Specifically,
we consider a dilute network of $N$ nodes, 
labelled $i=1\dots N$.
Each node is connected randomly to $c$ neighbors
with the symmetric connectivity matrix 
$\cA_{ij}=1,0$ for connected
and unconnected node pairs respectively. 
Each node $i$ has initial resource or {\it capacity} $\Lambda_i$,
randomly drawn from a distribution of $\rho(\Lambda_i)$.
Positive and negative values of $\Lambda_i$ correspond to
surplus and shortage of resources respectively.
These resources can be transported through connected links.
We denote by $y_{ij}\equiv -y_{ji}$ the current of resources 
from node $j$ to node $i$. 
The system minimizes the cost function
\begin{equation}
\label{eq_energy}
	 E=\frac{u^2}{2}\sum_i \Theta(-\xi_i)
	 +\sum_{(ij)}\frac{y_{ij}^2}{2}.
\end{equation}
$\xi_i\equiv \Lambda_i + \sum_j \cA_{ij}y_{ij}$ 
is the final resources of node $i$, 
and $\Theta(x)=1$ when $x>0$, 
and $0$ otherwise. 
In the first term, each unsatisfied node 
yields a shortage cost of $u^2/2$. 
The second term is the transportation cost of resources.

We first look for phase transitions in the model by numerical simulations.
To formulate an algorithm,
we introduce a variable $s_i=\pm 1$ for each node $i$ and
restrict its resources by $s_i \xi_i\ge 0$.
Introducing Lagrange multipliers $\mu_i$ for the resource constraint, 
we minimize the Lagrangian
\begin{equation}
\label{eq_lagr}
	L=
	\frac{u^2}{2}\sum_i\frac{1-s_i}{2}+\sum_{(ij)}\frac{y_{ij}^2}{2}
	+\sum_i\mu_i s_i \xi_i
\end{equation}
with the K\"uhn-Tucker conditions $\mu_i s_i \xi_i=0$
and $\mu_i\le 0$.
Optimizing $L$ with respect to $y_{ij}$,
one obtains
$y_{ij}=\mu_j s_j-\mu_i s_i$ and
$\mu_i =\min[0, (\Lambda_i+\sum_j \cA_{ij}\mu_j s_j)/s_i c]$.
Given a particular set of $\{s_i\}$, 
we iterate these equations to find the corresponding set of $\{\mu_i\}$.
The set of optimal $\{s_i\}$ is found by an approach similar to the
the {\it GSAT} algorithm \cite{selman1996},
by comparing the Lagrangian in \req{eq_lagr} for each choice of $\{s_i\}$.

%%%%%%%%% Figure1 %%%%%%%%%%%%%%%%%
\begin{figure}
\centerline{\epsfig{figure=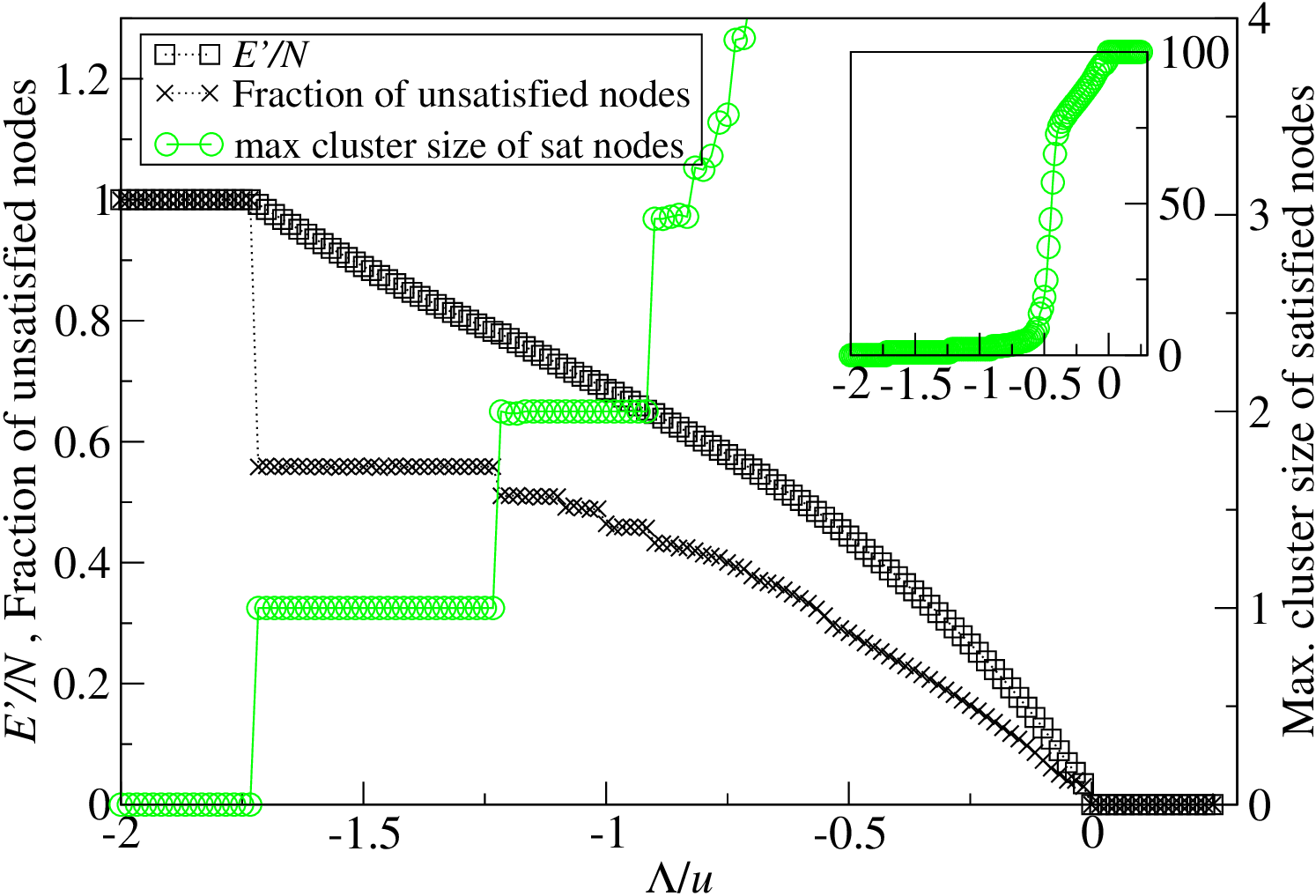, width=\linewidth}}
\caption{(Color online) Simulation results of average energy per node, 
the fraction of unsatisfied nodes
and the maximum cluster size of satisfied nodes after optimization,
as a function of $\Lambda/u$.
$E'=2E/u^2$.
Parameters:  $c=3$, $N=100$, 100 samples
and 1000 flips.
Inset: the maximum cluster size of satisfied nodes shown 
with a larger vertical scale.
}
\label{gr_energy}
\vspace{-0.5cm}
\end{figure}
%%%%%%%%%%%%%%%%%%%%%%%%%%%%%%%

We first consider networks with uniform capacity 
($\Lambda_i=\Lambda$ for all $i$).
As shown in Fig.~\ref{gr_energy} for $c=3$,
three phases can be identified:
(1) {\it unsatisfied} phase for $\Lambda/u\le -\sqrt{3}$,
in which all the nodes are unsatisfied and $E/N=u^2/2$;
(2) {\it partially satisfied} phase for $-\sqrt{3}<\Lambda/u<0$,
in which only some nodes are satisfied, 
and $0<E/N<u^2/2$;
(3) {\it satisfied} phase for $\Lambda/u\ge 0$,
in which all nodes are satisfied and $E/N=0$.

Unlike the relatively smooth decreasing trend of energy,
the fraction of unsatisfied nodes is a discontinuous function of $\Lambda/u$,
showing abrupt jumps at threshold values of $\Lambda/u$.
The step size of the curve decreases as $\Lambda/u$ increases,
and gradually becomes unresolvable by the numerical experiments.
This resembles the {\it Devil's staircase} observed in the circle map 
and other dynamical systems \cite{devaney1989}.
These threshold values of $\Lambda/u$ mark the position at which 
certain configurations of the satisfied nodes become energetically stable.

%%%%%%%%% Figure2 %%%%%%%%%%%%%%%%%
\begin{figure}
\centerline{\epsfig{figure=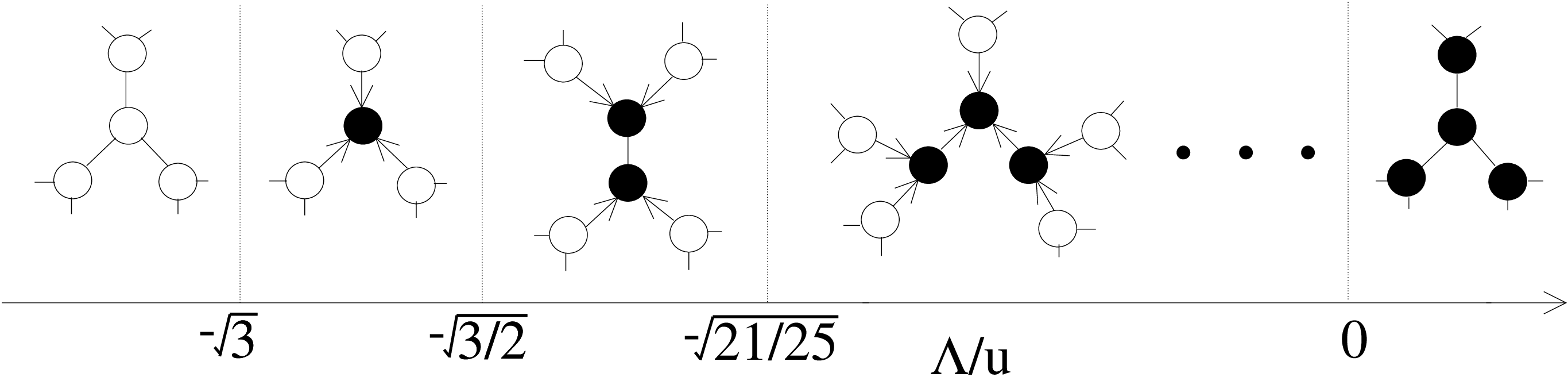, width=\linewidth}}
\caption{
The onset of different types of clusters of satisfied nodes for $c=3$,
with filled and unfilled circles 
representing satisfied and unsatisfied nodes respectively.
}
\label{mode}
\vspace{-0.5cm}
\end{figure}
%%%%%%%%%%%%%%%%%%%%%%%%%%%%%%%

To further confirm this picture,
we measure the average maximum cluster size of satisfied nodes in the samples.
Abrupt jumps of the cluster size 
are observed at the same threshold values as shown in Fig.~\ref{gr_energy}.
This indicates that new types of clusters of satisfied nodes are formed at
each jump,
as shown in Fig.~\ref{mode} for $c=3$.
The situation is similar to the cascades of phase transitions in RFIM 
due to the onsets of different clusters~\cite{bruinsma1983}. 
As shown in the inset of Fig.~\ref{gr_energy}, 
the maximum cluster size increases drastically to $O(N)$ around 
$\Lambda/u\approx -0.5$.
This corresponds to a percolation-like transition at which 
isolated clusters of satisfied nodes become connected.

We carry out the analysis using the Bethe approximation,
since the networks have a tree-like structure locally.
Messages are passed from the vertices of the sub-trees to the next layer,
resulting in a recursion relation of the messages.
In the cavity approach, 
these messages are the {\it cavity energy functions}, 
denoted by $E_j(y_{ij})$ for the energy of the tree 
terminated at a link from vertex $j$ to its ancestor $i$, 
when a current $y_{ij}$ is drawn from $j$ to $i$~\cite{wong2006}.
They can be expressed in terms of the energies of its descendents
$k=1,\dots, c-1$,
\begin{equation}
\label{cavity_energy}
	E_j(y_{ij})=
	\min_{\{y_{jk}\}}\Biggl[\sum_{k=1}^{c-1}E_k(y_{jk})
	+\frac{u^2}{2}\Theta(-\xi_j(y_{ij}))+\frac{y_{ij}^2}{2}\Biggr],
\end{equation}
where $\xi_j(y_{ij})=\Lambda_j+\sum_k y_{jk}-y_{ij}$.

%%%%%%%%% Figure3 %%%%%%%%%%%%%%%%%
\begin{figure}
\includegraphics[width=0.23\linewidth]{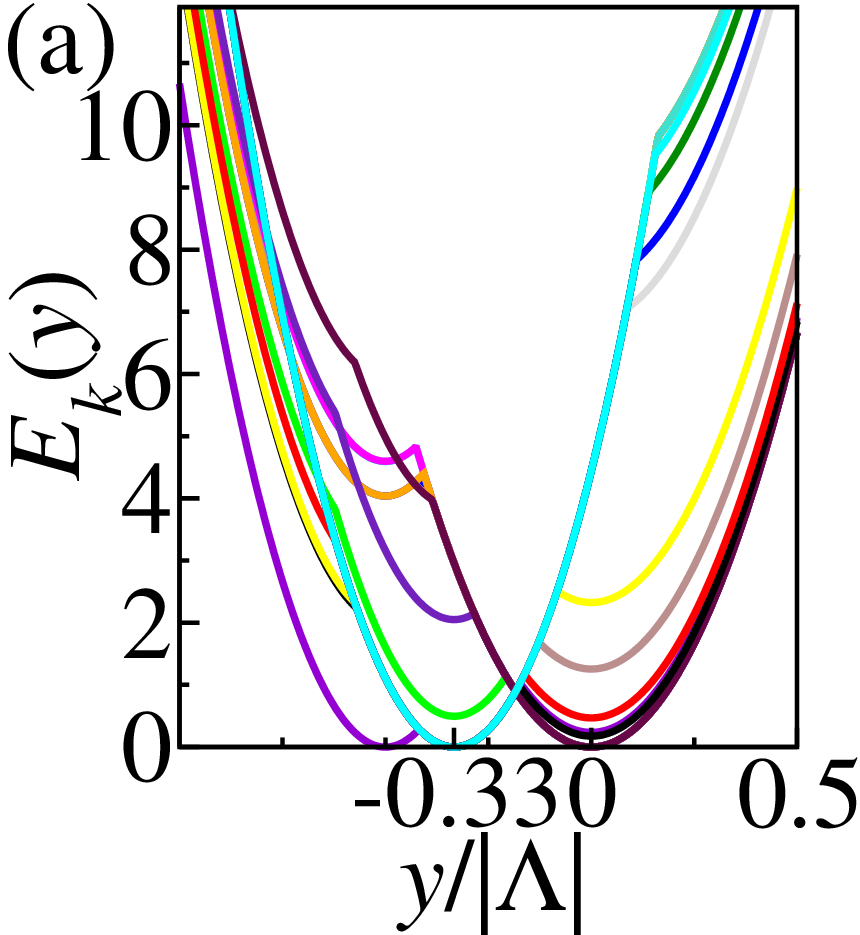}~
\includegraphics[width=0.23\linewidth]{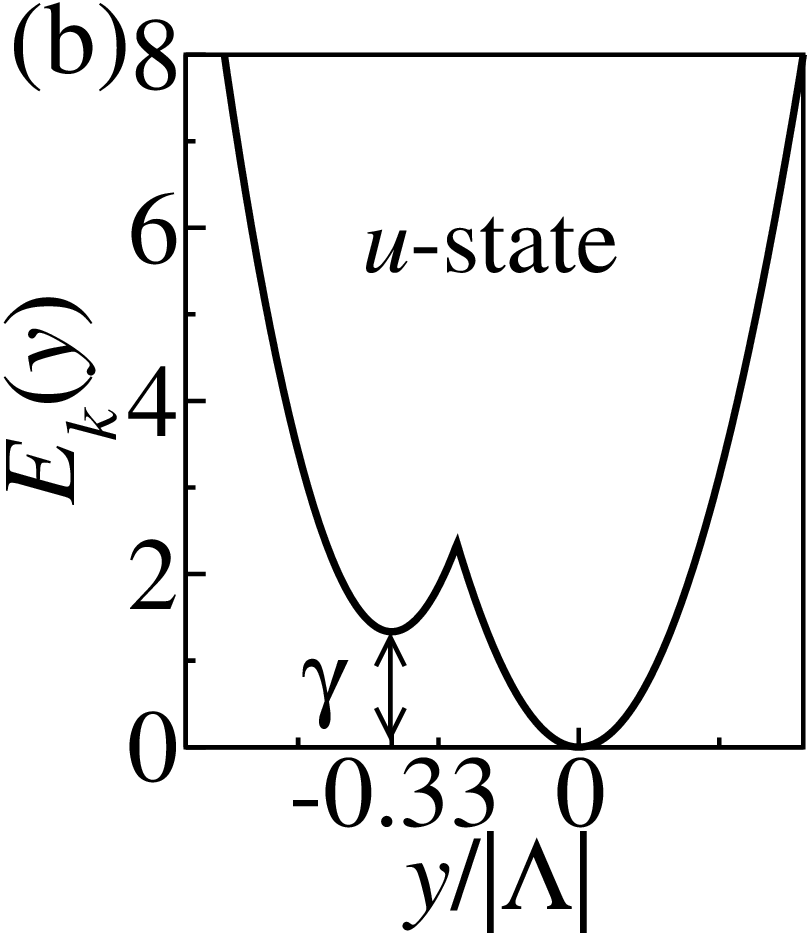}~
\includegraphics[width=0.23\linewidth]{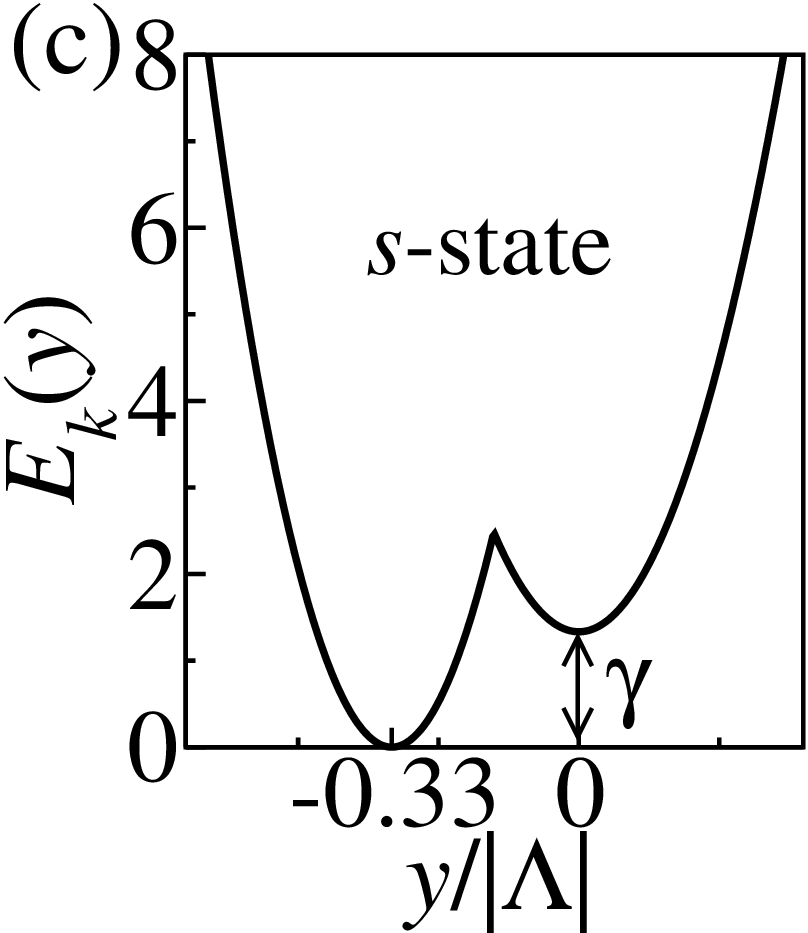}~
\includegraphics[width=0.23\linewidth]{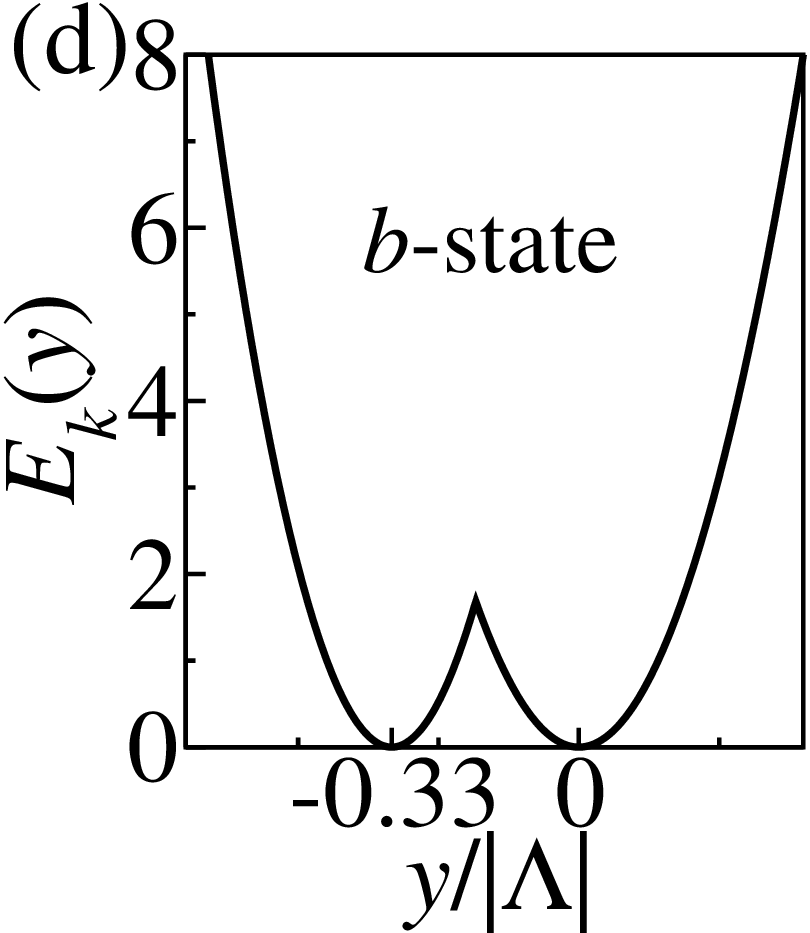}~
\caption{(Color online) (a) A typical set of cavity energy functions $E_k(y)$ 
at $c=3$, $\Lambda/u=-1.22 >-\sqrt{3/2}$.
(b-d) The closed set of $E_k(y)$ at $c=3$, 
$\Lambda/u=-5/3$ with
$\gamma=u^2/2-\Lambda/2c$,
corresponding to 
(b) the $u$-state,
(c) the $s$-state and
(d) the $b$-state.
}
\label{usb}
\end{figure}
%%%%%%%%%%%%%%%%%%%%%%%%%%%%%%%

In general,
the iteration of \req{cavity_energy} results in a distribution
of the cavity energy functions,
as shown in Fig.~\ref{usb}(a).
However,
in the regime $-\sqrt{c}\le\Lambda/u\le-\sqrt{c(c-1)/(c+1)}$,
it can be shown analytically that there are $c$ functional forms of $E_k(y)$
forming a closed set of solutions to the recursion relation 
in \req{cavity_energy}.
The three functions for $c=3$,
referred to as the $u$-, 
$s$- and $b$-states,
are shown in Fig.~\ref{usb}(b-d).
Respectively,
they correspond to states favoring unsatisfaction,
satisfaction, and bistability,
and have absolute minima at $y=0$,
$y=\Lambda/c$,
and both $y=0$ and $\Lambda/c$.
Furthermore,
numerical iterations of \req{cavity_energy} starting with random $E_k(y)$
show that this set of solutions is stable.
For $c>3$,
the closed set consists of more states having absolute minima at $y=0$,
and local minima at different energies at $y=\Lambda/c$,
similar to the $u$-state in Fig.~\ref{usb}.
As we shall see,
the closure property of the $c$ states greatly simplifies 
\req{cavity_energy}.
With the $u$- and $b$-states denoted as the $U$-state,
and the $s$-state by the $S$-state, 
their recursion relations are given by
an ``$SU$-recursion":
assign a vertex to an $S$-state 
if all its $c-1$ descendents are in the $U$-state, 
and to a $U$-state otherwise.

The recursion rules imply that optimization 
is achieved by the close packing of satisfied nodes in the network,
with the constraint that they do not form clusters.
Hence we call the regime $-\sqrt{c}<\Lambda/u<-\sqrt{c(c-1)/(c+1)}$
the {\it single-sat} regime.
This reduces the problem to the vertex cover problem~\cite{weigt2000}.
Since there is at most one satisfied node at the end of each link,
the maximization of the number of satisfied nodes is equivalent
to the minimization of the covered set size in the vertex cover problem.
Alternatively,
the model can be considered as the close packing limit of a lattice glass
model \cite{biroli2002}.
Both models 
exhibit glassy behavior,
and phases with multiple metastable states 
are found therein, 
corresponding to the computationally hard phases.

The description of the glassy behavior can be approached 
by the replica analysis~\cite{mezard1987}. 
In the so-called replica symmetric (RS) ansatz, 
one assumes that the network behavior is dominated by a single ground state. 
However, we find that this ansatz is not stable 
in the entire single-sat phase.
Instead, the network behavior is described 
by the so-called full replica symmetry-breaking (FRSB) ansatz, 
which assumes an infinite hierarchy of metastable states. 
Indeed, we have computed the optimized energy 
in the one-step RSB (1RSB) approximation, 
assuming that the optimized states with energy $E$ are distributed 
exponentially according to $\exp[N\Sigma(E)]$,
where $\Sigma(E)$ is the so-called configurational entropy~\cite{mezard2001}.
Futhermore, 
the 1RSB ansatz predicts that $\Sigma(E)$ exists up to an energy $E_d$
above the ground state energy $E_s$,
and the numerous metastable states prevent practical algorithms 
from converging to the true ground state,
resulting in a dynamical transition.
Specifically,
we find that in the single-sat regime,
$E/N=\Lambda^2/6+f_u(u^2/2-\Lambda^2/6)$,
where $f_u$ is the fraction of unsatisfied nodes.
In the RS approximation, 
$f_u=0.545$, 
whereas in the 1RSB approximation, 
$f_u=0.549$ and 0.550 at its $E_s$ and $E_d$ respectively.
Note that $f_u$ at the dynamical transition is close 
to the simulation result of $f_u=0.551$.

%%%%%%%%% Figure4 %%%%%%%%%%%%%%%%%
\begin{figure}
\includegraphics[width=0.3\linewidth]{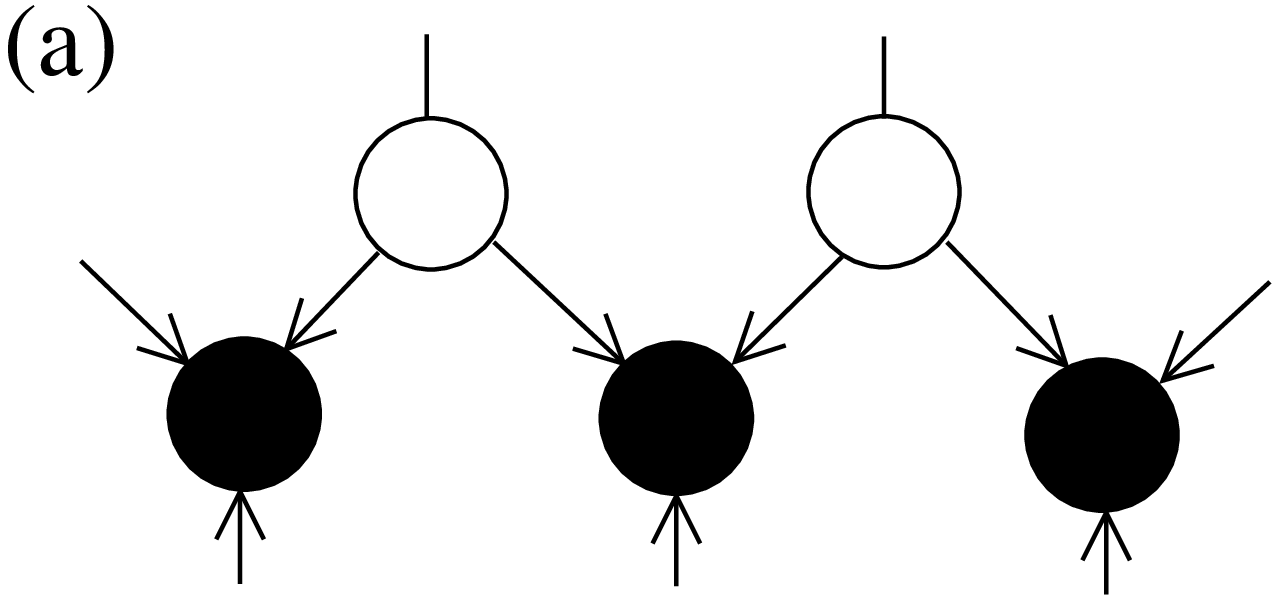}~
\hspace{1cm}
\includegraphics[width=0.3\linewidth]{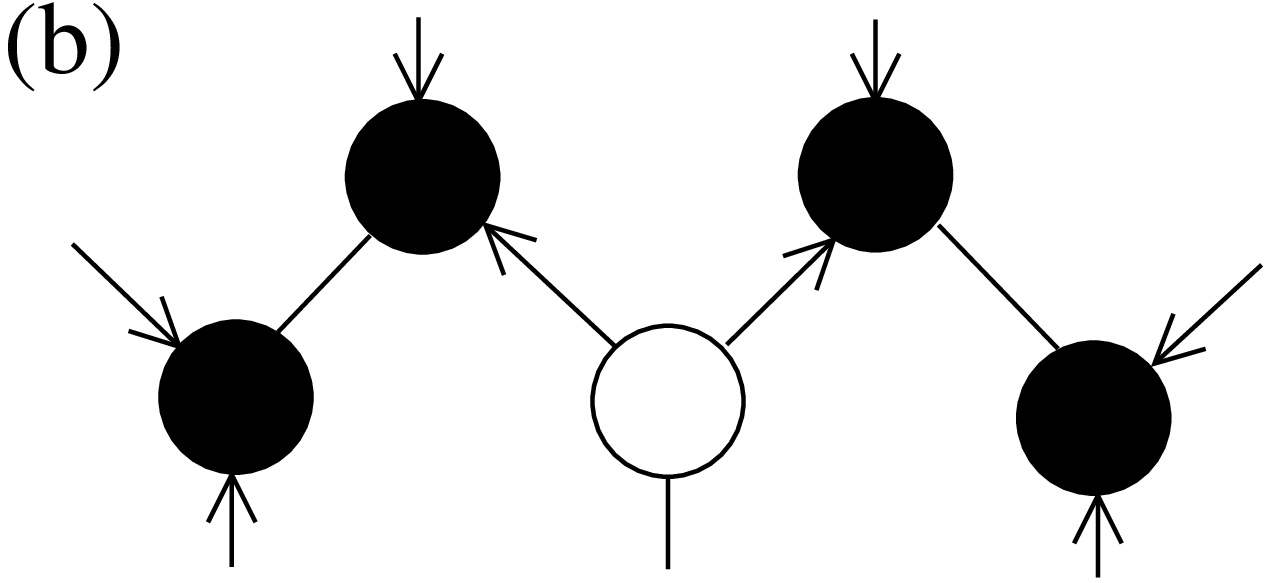}~
\caption{(a) A five-node configuration composing of 3 isolated 
satisfied nodes (filled circles).
It is stable when $\Lambda/u<-1$.
(b) A five-node configuration composing of 2 two-node clusters.
It replaces the configuration in (a) when $\Lambda/u>-1$.
}
\label{energyConfig}
\vspace{-0.5cm}
\end{figure}
%%%%%%%%%%%%%%%%%%%%%%%%%%%%%%%

When $\Lambda$ increases above the single-sat regime,
clusters of two satisfied nodes appear.
This {\it double-sat} regime exists in the range 
$-\sqrt{3/2}<\Lambda/u<-\sqrt{21/25}$ for $c=3$.
This is similar to the close packing limit of the Bethe glass model,
in which each occupied site can have at most one occupied neighbor.
However,
the present model is richer in behavior, 
as indicated by the jumps in the fraction of unsatisfied nodes 
in the double-sat regime of Fig.~\ref{gr_energy}.
They mark the positions at which configurations 
increasingly dominated by two-node clusters 
become energetically stable,
when $\Lambda$ increases.
An example of such an energy switch is shown in Fig.~\ref{energyConfig}.
Indeed,
such threshold values are expected at 
$\Lambda/u=-\sqrt{(3n-3)/(2n-1)}$ where $n=2,3,\dots$.
Only two of those thresholds are visible in Fig.~\ref{gr_energy},
probably due to the absence of configurations for larger $n$ 
in networks with $N=100$, 
and the limitations of the search algorithm.
Note that the double-sat regime may exhibit behaviors 
described by different RSB ansatz as $\Lambda/u$ varies.
It is also interesting to study the possible change of RSB behavior
as $\Lambda/u$ further increases up the Devil's staircase.

Next, 
we consider a simple case with quenched disorder of the node capacities.
In this case, 
$\Lambda_i$ is drawn from a bimodal distribution,
namely, $\Lambda_i=\Lambda_\pm$ with probabilities $f_\pm$,
where $\Lambda_-/u<-\sqrt{3}<\Lambda_+/u<-\sqrt{3/2}$ for $c=3$,
and $f_++f_-=1$.
$\Lambda_\pm$ are referred to as the rich and poor nodes respectively.
The recursion relations for the rich nodes 
follow the ``$SU$-recursion", 
whereas the poor nodes are always in the $U$-state. 
Note that the end points of this range are $f_+=0$ and 1,
corresponding to the unsatisfied and partially satisfied phases respectively.
Since these two phases are in the RS and RSB regimes respectively,
we expect that there is a phase transition when $f_+$ increases from 0 to 1.

Applying the RS ansatz to the region of low $f_+$,
we let $\puk$ be the probability that node $k$ is in the 
$U$-state.
Its recursion relation can be written as 
\begin{eqnarray}
\label{eq_recur1}
	p_{u,j}&&=\delta_{\Lambda_j, \Lambda_-}
	+\delta_{\Lambda_j, \Lambda_+}\biggr(1-\prod_{k=1}^{c-1}\puk\biggl),
\end{eqnarray}
which corresponds to an intense simplificaton of \req{cavity_energy}
In the easy regime at low $f_+$,
a stable fixed point solution with all $\puk=0$ or $1$ exists.
The site average $\avg{p_u}$ can thus be obtained from the equation 
$\avg{p_u}=f_-+f_+(1-\avg{p_u}^{c-1})$.
In terms of algorithms,
optimal network states in this regime can be obtained by 
the so-called belief propagation (BP) algorithm,
initializing the messages to 0 or 1.

The stability of the RS solution can be studied by considering
the propagation of fluctuations $\avg{(\delta p_{u,k})^2}$ under the 
recursion relation \req{eq_recur1} \cite{thouless1986}.
This leads to the Almeida-Thouless (AT) stability condition,
which reads
\begin{eqnarray}
\label{eq_AT}
	(c-1)f_+\avg{p_u^2}^{c-2}\le 1.
\end{eqnarray}
In the RS regime,
$\avg{p_{u}^2}=\avg{p_u}$ since $\puk=0$ or 1.
This implies an AT transition at $f_+^{\rm AT}=c^{c-2}/(c-1)^{c-1}$.

\begin{figure}
\centerline{\epsfig{figure=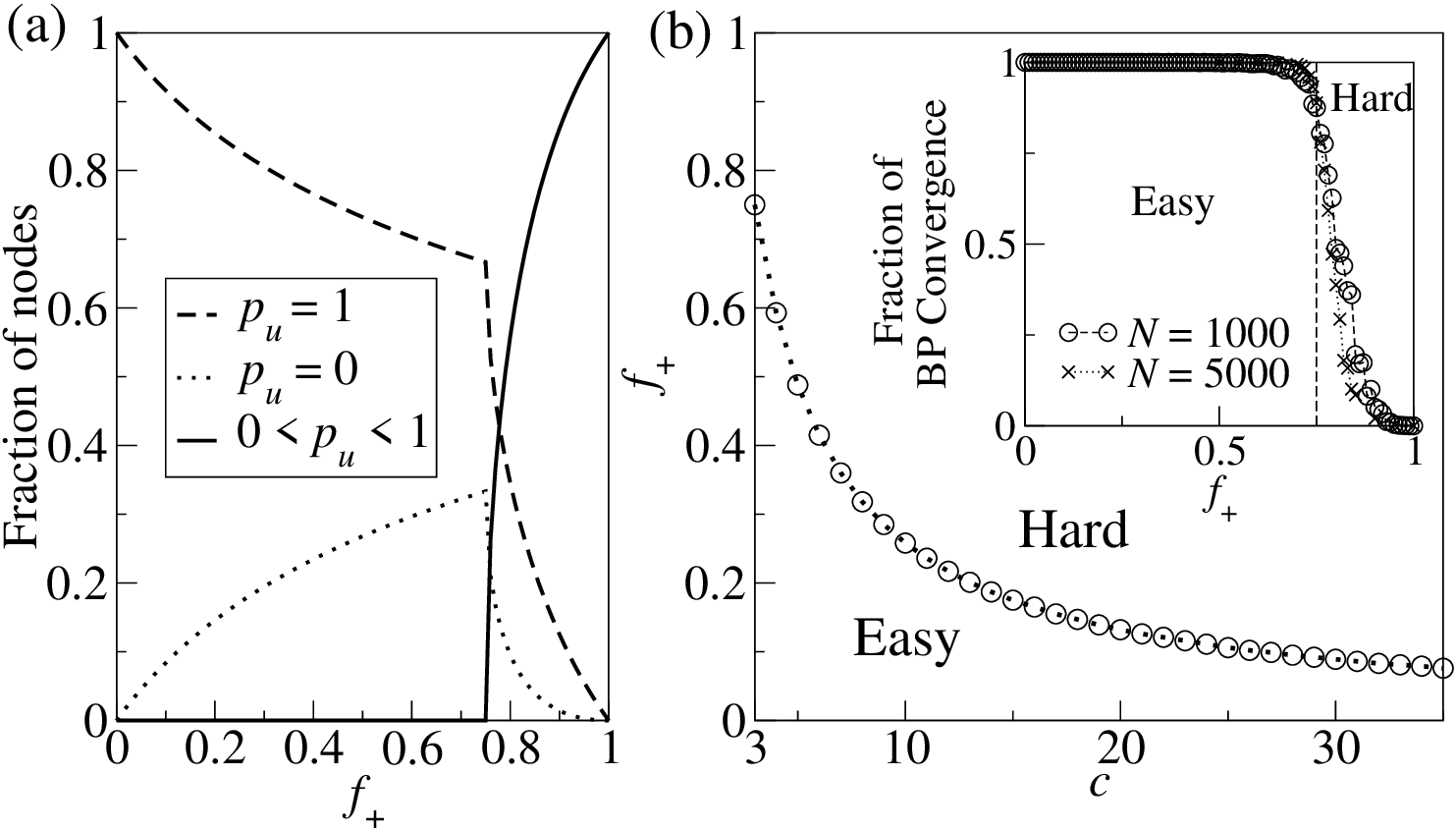, width=\linewidth}}
\vspace{-0.2cm}
\caption{(a) The fraction of nodes with 
$\puk=0$, 1, and $0<\puk<1$ for $c=3$.
(b) The phase diagram.
Inset: Simulation results of the fraction of messages from 
rich nodes with BP convergence on dilute networks
with $c=3$.
}
\label{solution}
\vspace{-0.4cm}
\end{figure}

As shown in Fig.~\ref{solution}(a), 
free nodes with $0<\puk<1$ start to exist 
when $f_+>f_+^{\rm AT}$,
analogous to the vertex cover~\cite{weigt2000} 
and graph coloring problems~\cite{mulet2002}.
This characterizes the hard region with multiple states.

%%%%%%%%% Figure5 %%%%%%%%%%%%%%%%%
%%%%%%%%%%%%%%%%%%%%%%%%%%%%%%%

As shown in the $f_+$-$c$ space 
in Fig.~\ref{solution}(b), 
the easy and hard phases exist 
below and above the AT line respectively.
In the large $c$ limit,
$f_+^{\rm AT}$ approaches $e/c$.
This result has an interesting connection with the vertex cover problem. 
Considering the cover set as the set of unsatisfied nodes, 
all links involving poor nodes are covered. 
The remaining links are those among the rich nodes, 
and the problem of minimizing the cover set size 
reduces to one that minimizes the subset size of covered nodes 
in the subnetwork of rich nodes. 
In the large $c$ limit, 
this subnetwork has a Poissonian connectivity distribution 
with an average $cf_+$. 
The result $cf_+^{\rm AT}=e$ agrees with the point of RS instability 
derived in~\cite{weigt2000}.

The AT transition has implications on the algorithms.
We consider the BP algorithm initialized with messages 0 and 1.
As shown in Fig.~\ref{solution}(b) inset,
effectively all messages from rich nodes converge in the easy regime.
However, 
a significant fraction of messages 
fluctuates between 0 and 1 above $f_+^{\rm AT}$,
indicating the breakdown of the RS ansatz.
Algorithmically,
decimation procedures, 
such as those used in the survey propagation (SP)
algorithm~\cite{mezard2002}, are required.

In summary, 
we have studied how nonlinearities affect the flow patterns 
in transportation networks. 
In the case of uniform capacity,
phase transitions resembling the Devil's staircase 
reveal the cascades of clustered flow patterns. 
In the single-sat regime 
with a closed set of only a few cavity energy functions, 
the flow pattern has a correspondence with the vertex cover problem.
Glassiness arises from the frustration in competitions for resources.
%Incorporating the picture of multiple metastable states typical of glasses,
%we obtain energy estimates with improved agreement with simulated results.
In the case with quenched disorder of the capacities, 
an increase in the fraction of rich nodes 
induces a phase transition from an easy phase to a hard one, 
in which message-passing algorithms experience convergence problems. 
These features are relevant to general network optimization problems 
with nonlinear elements.

We thank David Saad and Jack Raymond for meaningful discussions 
and Patrick Lee for encouragement. 
This work is supported
by the Research Grants Council of Hong Kong
(HKUST 603606, 603607 and 604008).

%%%%%%%%%%%%%%%%%%%%%%%%%%


\begin{thebibliography}{0}

\bibitem{doyle1984}
P. G. Doyle and J. L Snell, 
{\it Random Walks and Electric Networks} 
(Mathematical Association of America, Washington D. C., 1984).

\bibitem{rockafellar1984}
R. T. Rockafellar, 
{\it Network Flows and Monotropic Optimization} 
(Wiley, New York, 1984).

\bibitem{banavar2000}
J. R. Banavar, F. Colaiori, A. Flammini, A. Maritan, and A. Rinaldo, 
Phys. Rev. lett. {\bf 84}, 4745 (2000); 
M. Durand, 
Phys. Rev. Lett. {\bf 98}, 088701 (2007); 
S. Bohn and M. O. Magnasco, 
Phys. Rev. Lett. {\bf 98}, 088702 (2007);
Z. Shao and H. Zhou, 
Phys. Rev. E {\bf 75}, 066112 (2007).

\bibitem{imry1975}
Y. Imry and S. K. Ma, 
Phys. Rev. Lett. {\bf 35}, 1399 (1975).

\bibitem{bruinsma1983}
R. Bruinsma and G. Aeppli, 
Phys. Rev. Lett. {\bf 50}, 1494 (1983); 
R. Bruinsma, Phys. Rev. B {\bf 30}, 289 (1984).

\bibitem{toulouse1977}
G. Toulouse, 
Commum. Phys. {\bf 2}, 115 (1977).

\bibitem{mezard2002}
M. M\'ezard and R. Zecchina,
Phys. Rev. E {\bf 66}, 056126 (2002).

\bibitem{mulet2002}
R. Mulet, A. Pagnani, M. Weigt, and R. Zecchina,
Phys. Rev. Lett. {\bf 89}, 268701 (2002); 
A. Braunstein, R. Mulet, A. Pagnani, M. Weigt, and R. Zecchina,
Phys. Rev. E {\bf 68}, 036702 (2003).

\bibitem{nishimori2001} 
H.~Nishimori, 
{\it Statistical Physics of Spin Glasses and Information Processing}
(Oxford University Press, Oxford, UK, 2001).

\bibitem{mezard1987}
M. M\'ezard, G. Parisi and M. A. Virasoro, 
{\it Spin Glass Theory and Beyond}
(World Scientific, 1987).

\bibitem{wong2006} 
K.~Y~.M.~Wong and D.~Saad, 
Phys. Rev. E {\bf 74}, 010104 (2006); 
Phys. Rev. E {\bf 76}, 011115 (2007).

\bibitem{shenker1996}
S. Shenker, D. Clark, D. Estrin and S. Herzog,
Comput. Commun. Rev. {\bf 26}. 19 (1996); 
R.~L. Rardin
{\it Optimization in Operations Research}
(Prentice Hall, Englewood Cliffs, NJ, 1998).

\bibitem{garey1979}
M. R. Garey and D. S. Johnson, 
{\it Computers and Intractability: A Guide to the Theory of NP-Completeness} 
(Freeman, New York, 1979).

\bibitem{weigt2000}
M. Weigt and A.~K. Hartmann,
Phys. Rev. Lett. {\bf 84}, 6118 (2000); 
Phys. Rev. E {\bf 63}, 056127 (2001).

\bibitem{selman1996}
B. Selman, H. Kautz and B. Cohen
{\it DIMACS Series in Discrete Mathematics and 
Theoretical Computer Science} {\bf 26}, 521 (1996).

\bibitem{devaney1989}
R.~L. Devaney, 
{\it An Introduction to Chaotic Dynamical Systems}
(Addison-Wesley, Redwood City, CA, 1989).

\bibitem{biroli2002}
G. Biroli and M. M\'ezard,
Phys. Rev. Lett. {\bf 88}, 025501 (2001);
O. Rivoire, G. Biroli, O.~C. Martin, and M. M\'ezard, 
Eur. Phys. J. B {\bf 37}, 55 (2004).

\bibitem{mezard2001}
M. M\'ezard and G. Parisi,
Eur. Phys. J. B {\bf 20}, 217 (2001).

\bibitem{thouless1986}
D. J. Thouless,
Phys. Rev. Lett. {\bf 56}, 1082 (1986).

\end{thebibliography}
\end{document}